# Multilayer black phosphorus as broadband saturable absorber for pulsed lasers from 1 to 2.7 μm wavelength

Lingchen Kong,[1] Zhipeng Qin,[1] Guoqiang Xie,[1*] Zhinan Guo,[2] Han Zhang,[2*] Peng Yuan,[1] and Liejia Qian[1]

[1]Key Laboratory for Laser Plasmas (Ministry of Education), IFSA Collaborative Innovation Center, Department of Physics and Astronomy, Shanghai Jiao Tong University, Shanghai, 200240, China,
[2]SZU-NUS Collaborative Innovation Centre for Optoelectronic Science & Technology, and Key Laboratory of Optoelectronic Devices and Systems of Ministry of Education and Guangdong Province, College of Optoelectronic Engineering, Shenzhen University, Shenzhen 518060, China.

**Abstract**

It attracts wide interest to seek universe saturable absorber covering wavelengths from near infrared to mid-infrared band. Multilayer black phosphorus, with variable direct bandgap (0.3-2 eV) depending on the layer number, becomes a good alternative as a universe saturable absorber for pulsed lasers. In this contribution, we first experimentally demonstrated broadband saturable absorption of multilayer black phosphorus from 1 μm to 2.7 μm wavelength. With the as-fabricated black phosphorus nanoflakes as saturable absorber, stable Q-switching operation of bulk lasers at 1.03 μm, 1.93 μm, 2.72 μm were realized, respectively. In contrast with large-bandgap semiconducting transition metal dichalcogenides, such as $MoS_2$, $MoSe_2$, multilayer black phosphorus shows particular advantage at the long wavelength regime thanks to its narrow direct bandgap. This work will open promising optoelectronic applications of black phosphorus in mid-infrared spectral region and further demonstrate that BP may fill the gap of between zero-bandgap graphene and large-bandgap TMDs.

Introduction

Black phosphorus (BP), a layered allotrope of phosphorus, owns its unique properties. Due to puckered layer or zigzag direction of intralayer atoms, the BP crystal exhibits highly anisotropic mechanical properties [1]. Moreover, few-layered BP presents highly anisotropic electric conductance and strain-controlled anisotropic electric mobility [2]. By use of its high carrier mobility, few-layer BP has been manufactured into fast field-effect transistors (FETs) and opens its electronic applications [3-4]. Similar to $MoS_2$, BP also possesses thickness-depended energy bandgap with 1.5-2.0 eV for single layer [5-7], 0.59 eV for five layers [8] and 0.3-0.33 eV for bulk sample [1, 3]. However, $MoS_2$ shows a transition from direct bandgap to indirect bandgap once it becomes multilayer while BPs always show direct bandgap property [5]. Such difference would endow multilayer BPs with intrinsically stronger light-matter interaction than multilayer $MoS_2$. This might lead to some unique applications of multilayer BPs for optoelectronics. Meanwhile, the narrow direct bandgap of multilayer BP results in strong absorption of photons at mid-infrared region. Recently multilayer BP with controllable size, acquired by liquid phase exfoliation [9], has been verified to possess optical saturable absorption at 1.5 μm wavelength [10]. This reminds us to fabricate a universe saturable absorber covering from near infrared to mid-infrared band with multilayer BPs due to their narrow direct bandgap.

Up to now, saturable absorbers based on two dimensional (2D) materials such as graphene [11-13], topological insulators (TIs) [14-16], semiconducting transition metal dichalcogenides (TMDs) [17-18] have been rapidly developed for pulsed lasers from 1 μm to 2 μm wavelength. But there always exists bottleneck for saturable absorbers at the longer wavelength of ~3 μm. The zero-bandgap structure of graphene weakens the absorption at longer wavelength, while the large bandgap of TMDs prevents absorber from photon absorption at longer wavelength. Fortunately, multilayer BP, with direct intrinsic bandgap as narrow as ~0.3 eV for bulk sample [19, 20], should be a suitable candidate for saturable absorber at mid-infrared region. So far, saturable absorption of 2D materials at ~3 μm wavelength has not been verified before. By incorporating multilayer BP-based passive optical modulating device inside solid-state lasers at different wavelengths, we are able to verify the broadband saturable absorption of BP. In this contribution, we demonstrate that BP can operate as a broadband saturable absorber from 1 μm to 2.7 μm wavelength. By adopting multilayer BP-coated gold-film mirror as optical modulating devices in the oscillators, Q-switched lasers at wavelengths of 1.03 μm, 1.94 μm and 2.72 μm were demonstrated, respectively. To the best of our knowledge, this is the widest passive Q-switcher based on two-dimensional materials.

**Material characterization and experimental setup**

The multilayer BP sample used in the experiments was fabricated by liquid phase exfoliation (LPE) method with N-methylpyrrolidone (NMP) as solvent [10, 21]. For Raman measurement, TEM and ATM test, the sample was prepared by drop-casting the BP dispersion onto the surface of the quartz substrate and then dried in a vacuum drying oven. Raman spectrum of the BP flakes is shown in figure 1(a). From the Raman spectrum, it can be seen that there are three Raman peaks located at 361.9 cm$^{-1}$, 438.1 cm$^{-1}$ and 465.8 cm$^{-1}$, which are corresponding to the BP's out-of-plane vibration mode $A_g^1$, in-plane vibration modes $B_{2g}$ and $A_g^2$, respectively [22]. To characterize the morphology and thickness of the multilayer BP sample, transmission electron microscope (TEM) and atomic force microscopy (AFM) measurements were performed. Figure 1(b) shows the TEM image in micron scale. It can be seen that the exfoliated BP is in the form of layered structure with around 4 micrometer in diameter. The thickness parameter of the BP flake is shown in figure 1(c) and figure 1(d). AFM image result shows that the thickness of the BP flake is around 23 nm, indicating this type of BP possesses comparable bandgap of bulk BP with photon absorption at wavelength as far as 4.1 μm.

For fabricating a broadband high-reflective saturable absorber as passive optical modulator, we dropped the BP dispersion onto a gold-film mirror and dried it. Then the BP-coated mirrors were characterized and used for Q-switched lasers. The inset of figure 2(a) shows the photograph of the BP gold-film mirror, and multilayer BP can be clearly seen on the mirror. To verify broadband absorption of the BP sample, reflectivity spectrum was measured by using a Lambda 950 PerkinElmer UV/VIS/NIR spectrometer with result shown in figure 2(a). It can be seen that BP sample shows a strong absorption at the whole spectrum band from ~0.5 to 2.5 μm wavelength, implying BP's ultrabroadband absorption characteristics.

To further characterize the saturable absorption characteristics of BP-coated gold film mirror, we perform the saturable absorption measurement with a self-built picosecond fiber laser at 1.94 μm wavelength. The measurement result is shown in figure 2(b). Considering a Gaussian beam of incident light, the effective reflectivity R of the saturable absorber mirror satisfies:

$$R(F) = 1 - \alpha_{ns} - \Delta R \left[1 - \frac{F_{sat}}{F} \ln\left(1 + \frac{F}{F_{sat}}\right)\right] \qquad (1)$$

where $F$ is incident light fluence, $\alpha_{ns}$ is linear loss by absorber and substrate, $\Delta R$ is modulation depth, and $F_{sat}$ is saturation fluence of absorber. Fitting with Eq. 1, the modulation depth $\Delta R$ and saturation fluence $F_{sat}$ of the BP gold-film mirror are determined to be 7.8% and 1.15 μJ/cm$^2$, respectively. The linear loss portion $\alpha_{ns}$ from BP and gold-film substrate is 20.5%. The large modulation depth and the relatively low saturation fluence indicate that the BP gold-film mirror is suitable as Q-switcher of bulk lasers.

The schematic of the laser cavity used for the three Q-switched lasers was shown in figure 3. Laser diode was used as the pumping source. The pump beam was collimated and focused into gain media by two coupling convex lenses L1 and L2. The three plano-concave mirrors M1, M2, and M3 had the same radius of curvature of -100 mm, and were all highly reflectively coated for laser wavelength and anti-reflectively coated for pump wavelength. The output coupling was 2% for 1.03 μm and 1.93 μm wavelength, and 1% for 2.72 μm wavelength. For 1.03 μm laser, the pump source was a fiber-coupled laser diode at 940 nm and the gain medium was Yb:LuYAG crystal. For 1.93 μm laser, the pump source was a single-emitter laser diode at 790 nm and the gain medium was Tm:CaYAlO$_4$ crystal. The 2.72 μm laser was pumped by fiber-coupled laser diode at 976 nm with Er:Y$_2$O$_3$ ceramic as gain medium. The multilayer BP saturable absorber (BP-SA) was used as an end mirror in the cavity.

**Results and discussions**

In order to further verify the broadband saturable absorption of BP sample, we conducted the Q-switching experiment in bulk lasers at 1.03 μm, 1.93 μm and 2.72 μm wavelengths, respectively. A general X-folded cavity was employed to operate the Q-switched lasers. Description in the method section gives the details of the cavity setup. Using BP gold-film mirror as an end mirror, we realized stable Q-switched pulses at 1.03 μm, 1.93 μm and 2.72 μm, respectively. The typical Q-switched pulse trains, pulse durations and the laser spectra at three wavelengths are summarized in figure 4.

Firstly, we tested the Q-switching performance of BP gold-film mirror in Yb:LuYAG bulk laser at 1 μm wavelength. Under an incident pumping power of 4.0 W, the stable Q-switched pulses was achieved with average output power of 6 mW, pulse width of 1.73 μs, repetition rate of 63.9 KHz, and pulse energy of 0.09 μJ at the wavelength of 1029 nm. The low output power was mainly attributed to the large linear loss of the BP gold film mirror.

The Q-switching performance of the BP mirror was also tested in Tm: CaYAlO$_4$ bulk laser at 2 μm wavelength. With an incident pumping power of 5.1 W, the laser emitted stable Q-switched pulses with average output power of 12 mW, pulse repetition rate of 17.7 KHz, pulse width of 3.1 μs, and pulse energy of 0.68 μJ at 1930 nm. The obtained pulse energy and pulse duration are comparable to the 2 μm Q-switching results with graphene as saturable absorber [23].

To verify saturable absorption of multilayer BP at longer wavelength, Q-switching experiment at 2.7 μm wavelength was also implemented in Er:Y$_2$O$_3$ ceramic laser. Due to a long lower-level lifetime of Er:Y$_2$O$_3$ and a large vapor absorption loss in atmosphere, Er:Y$_2$O$_3$ ceramic laser at 2.72 μm wavelength generally has high laser threshold. For reducing the loss induced by the BP mirror, in the experiment we adopted a BP gold film mirror with less BP layers, which had a reflectivity of ~90%. With the thin BP mirror as saturable absorber, Q-switching operation at 2.72 μm wavelength was realized with average output power of 6 mW, pulse width of 4.47 μs, pulse repetition rate of 12.6 kHz and pulse energy of 0.48 μJ. The results demonstrate that BP is a promising saturable absorber as passive optical modulator in mid-infrared spectral band.

**Conclusions**

In summary, we have experimentally demonstrated broadband saturable absorption of BP from 1 μm to 2.7 μm wavelength. The saturable absorption measurement at 1.94 μm demonstrates the as-fabricated BP gold-film mirror has a modulation depth of 7.8% and saturation fluence of 1.15 μJ/cm$^2$. With the BP sample as saturable absorber, Q-switched lasers at 1.03 μm, 1.93 μm and 2.72 μm by using different gain media have been realized. The results show that BP is potential as a universe saturable absorber with ultrawide absorption band. Considering the bandgap (~0.3 eV) of multilayer BP, saturable absorption at 2.72 μm wavelength are approaching its absorption limit at mid-infrared region. The easy fabrication, broadband absorption and low saturation fluence of multilayer BP would promote optoelectronic applications towards the unexplored mid-infrared region.

**Acknowledgment**

We would be grateful to Prof. Yuzhi Zhang and Prof. Lingnan Wu in Shanghai Institute of Ceramics for measuring the reflection spectrum of the BP gold-film mirror. We would also be grateful to Prof. Jian Zhang and Prof. Xiaodong Xu in Jiangsu Normal University for providing the laser crystals. The work is partially supported by the Excellent Academic Leader Project of Shanghai (Grant No. 15XD1502100), National Natural Science Foundation of China (Grant No. 11421064 and 61435010) and the National Basic Research Program of China (Grant No. 2013CBA01505).

**Figures**

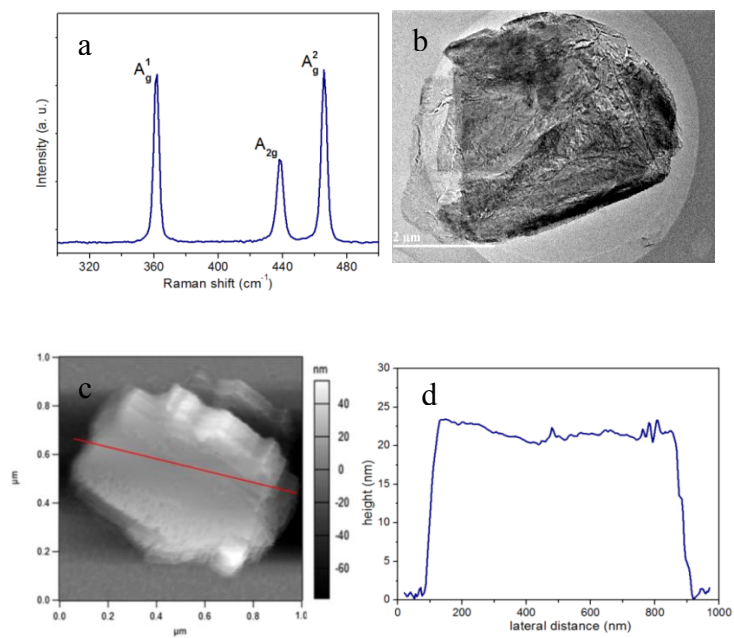

**Figure 1**. (a) Raman spectrum of the multi-layer BP nanoflake; (b) TEM image of the multi-layer BP nanoflake; (c) 2D graphical AFM image of BP nanoflake; (d) Height profile of the section marked in (c).

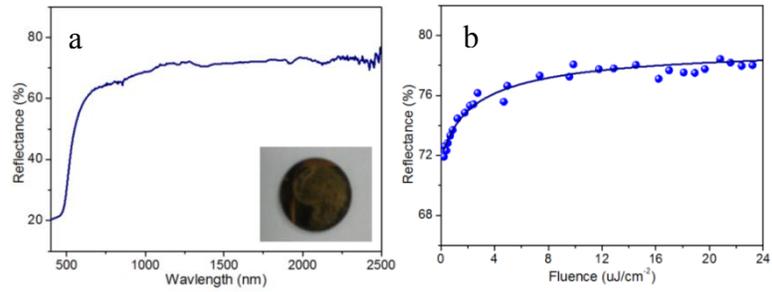

**Figure 2**. (a) The reflectivity spectrum of multilayer BP gold-film mirror. Inset of (a) is photograph of BP multilayer gold-film mirror. (b) The power-dependent nonlinear reflectivity of the multilayer BP gold-film mirror.

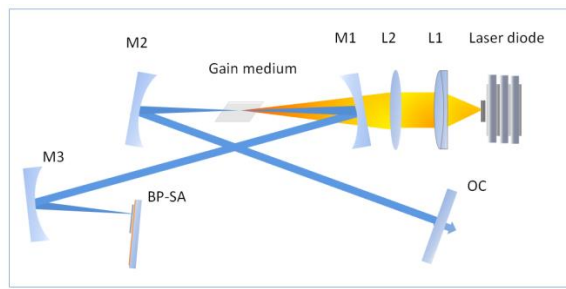

**Figure 3**. Schematic of Q-switched laser cavity with multilayer BP as saturable absorber.

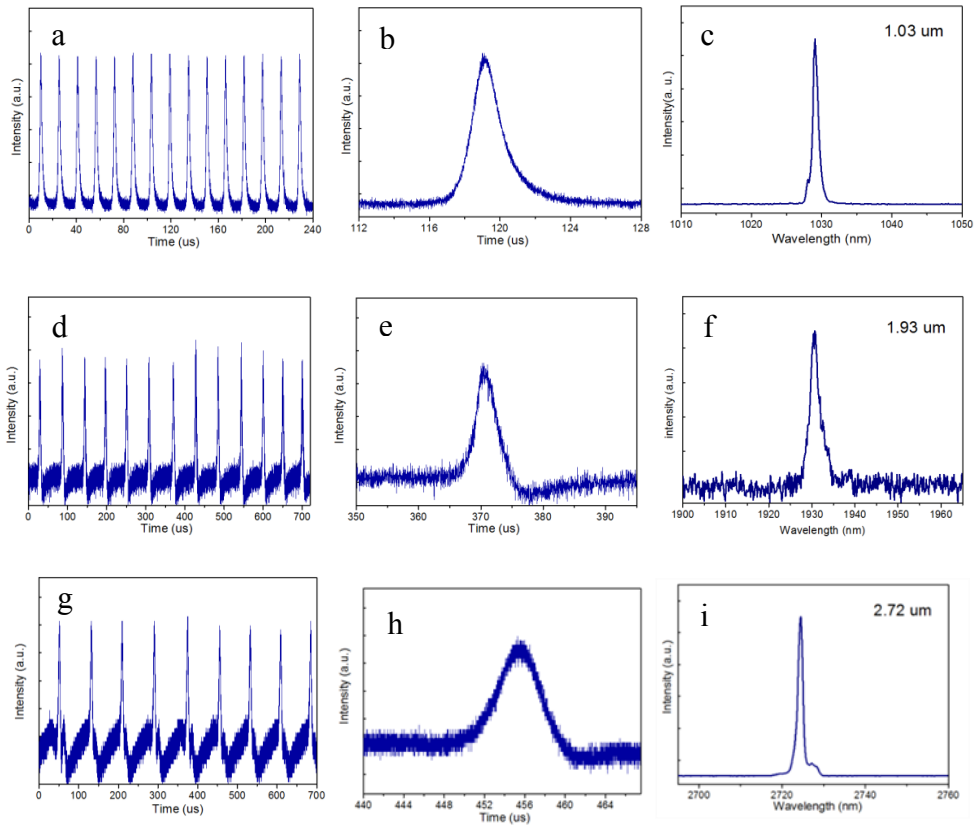

Figure 4. Q-switching pulse results based on multi-layer BP gold-film mirror at wavelengths of 1.03 μm, 1.93 μm and 2.72 μm, respectively. (a), (b) and (c) are typical pulse train, pulse profile and laser spectrum from 1.03 μm Yb:LuYAG laser, respectively; (d), (e) and (f) are those of Q-switched 1.93 μm Tm:CaYAlO$_4$ laser; (g), (h) and (i) are those of Q-switched 2.72 μm Er:Y$_2$O$_3$ laser.